\documentclass[12pt]{article}

\usepackage{vmargin}   
\setmarginsrb{3cm}{2cm}{3cm}{2cm}{1cm}{1cm}{1cm}{1cm}   
\usepackage{amsmath}   
\usepackage{amssymb}  
\usepackage[dvips]{graphicx}  
\usepackage{rotating}   
\usepackage{psfrag} 
\usepackage{fancyhdr}

\author{Karim Noui\footnote{noui@lmpt.univ-tours.fr} \\
Laboratoire de Math\'ematiques et de Physique Th\'eorique \\
UMR/CNRS 6083, F\'ed\'eration Denis Poisson \\
Facult\'e des Sciences et Techniques\\
Parc de Grandmont, 37200 Tours, France }
\title{\bf A model for the motion of a particle \\
in a quantum background}
\date{} 

\begin{document} 
\sloppy
\maketitle  

\begin{abstract}
We are studying the dynamics of a one-dimensional field in a non-commutative Euclidean space.
The non-commutative space we consider is the one that
emerges in the context of three dimensional Euclidean quantum gravity: it is a deformation 
of the classical Euclidean space $\mathbb E^3$ and the Planck length $\ell_P$ plays the role
of the deformation parameter.
The field is interpreted as a particle which evolves in a quantum background. 
When the dynamics of the particle is linear, the resulting motion is similar to the 
standard motion in the classical space $\mathbb E^3$. However, non-linear dynamics on the non-commutative
space are different from the corresponding non-linear dynamics on the classical space. 
These discrepencies are interpreted as ``quantum gravity" effects. 
Finally, we propose a background independent description of the propagation of the particle in the quantum
geometry.
\end{abstract}

\newpage

\section{Motivations}
Loop Quantum Gravity (LQG) \cite{LQG} has been one of the first theory to turn the question of space-time
structure at the Planck scale into a mathematically well-posed problem. This is surely one of the most important and most
beautifull achievement of LQG. Indeed, LQG provides a background independent quantization of
general relativity where standard geometrical notions, like length, area or volume, become well-defined operators
acting on a suitable Hilbert space \cite{Area}. 
Then, the problem of finding how looks space-time at the Planck scale turns out 
to be the problem of finding the eigeinvalues of these operators. Even if the answer to this last question is still
controversary for different reasons (see the recent works (\cite{DT}) for instance), 
it has allowed the possibility to address many other fundamental issues (black hole thermodynamics \cite{BH}, questions of singularities in classical gravity \cite{singularities}) 
that we sometimes did not even know how to tackle before. Even if some of these results
are still under discussion, one can claim that LQG offers a simple mathematical framework where
one can properly study fundamental aspects of quantum gravity. These last years, we have also seen the emergence of
a quantum gravity phenomenology \cite{DSR} where models have been proposed to describe the low energy regime of quantum
gravity. It is nonetheless important to underline that these models, which exhibit very interesting effects, 
are strongly discussed in the literature and their link with LQG is not clear at all in four dimensions.

In three dimensions (Euclidean signature and no cosmological constant), 
the situation is somehow simpler: it was argued that
quantum gravity effects could be completely recasted into
non-commutative effects \cite{FL}. In that picture,
space-time would become non-commutative at the Planck scale and its ``isometry" algebra 
would be a deformation of the standard classical algebra, known as the Drinfeld double. 
The deformation parameter
is the Newton constant $G$ (or equivalently the Planck length $\ell_P=\hbar G$). Recently, it was precisely shown
that this non-commutative space-time, in the Euclidean regime, admits a fuzzy space representation \cite{Noui}. 
As a consequence, this model of three dimensional Euclidean quantum space-time has in fact a discrete 
structure at the Planck scale and the dynamics of fields evolving in such a space become discrete as well. 
The purpose of
this article is to illustrate the effects of the discreteness with some simple but enlighting examples. 
More precisely, we will consider the dynamics of a one-dimensional field (it depends only on
one coordinates out of the three) interpreted as the motion of
a particle in a quantum background. 
At this point, it is important to underline that 
the system we are studying is a model for the dynamics of particle in a quantum background
based on two main asumptions: 
$(i)$ the quantum background is assumed to be the one that admits
the Drinfeld double as its deformed ``isometry algebra"; 
$(ii)$ the one-dimensional field is interpreted as a particle.

The first Section is devoted to briefly recall the construction of the non-commutative space.
We start by underlining the importance of the quantum double (or Drinfeld double) $DSU(2)$
in that construction: the non-commutative space is indeed defined as the space that admits $DSU(2)$
as isometry algebra. As the space is non-commutative, it is indirectly described in terms
of its algebra of functions which appears to be the convolution algebra $C(SU(2))^*$ of distributions on the group
$SU(2)$. Using harmonic analysis, we show that $C(SU(2))^*$ is isomorphic to the space $\oplus_n \text{Mat}_n(\mathbb C)$
where $\text{Mat}_n(\mathbb C)$ are the set of complex matrices of dimension $n$: this makes clear that 
the non-commutative space is fuzzy. To make concrete that $C(SU(2))^*$ is a deformation of
the algebra $C(\mathbb E^3)$ of functions on $\mathbb E^3$,  we exhibit a link between these two spaces. More precisely,
in this article, we restrict our study to the space of functions $C(SU(2))$ and show that it is isomorphic to
the direct sum $C_{B_{\ell_P}}(\mathbb E^3)\oplus C_{B_{\ell_P}}(\mathbb E^3)$ where $C_{B_{\ell_P}}(\mathbb E^3)$
is the sub-space of $C(\mathbb E^3)$ of functions whose spectrum belongs to the open sphere $B_{\ell_P}$ of 
radius $\ell_P^{-1}$.
The general result for $C(SU(2))^*$ is given in \cite{Noui}. 
Thus, any function $\phi \in C(SU(2))$ can be equivalently described by a matrix $\hat{\Phi}$ 
or by a pair of continuous functions $\Phi_+ \oplus \Phi_-$.
We define an integral and derivative operators on $C(SU(2))^*$ which allows to write an action for a scalar field
on the non-commutative space. We finish the first Section with a study of the free action for the scalar field.
In the second Section, we focus on the dynamics of a one-dimensional field
 which is
interpreted as a particle evolving in a given potential 
along one ``time" direction: equations of motion are written, solutions are found and discussed. In particular, 
the field admits two components $\Phi_\pm(t)$ when written in the continuous representation:
if $\Phi_+$ described the motion of a particle then, in some generic cases, 
$\Phi_-$ is the backward motion in the sense that
$\Phi_+(t)=\Phi_-(-t)$. Thus, there is a kind of miror symmetry between the two components.
We show that, in the case of a (free) quadratic potentiel, solutions are similar to standard classical ones.
Important differences occur when one considers non-linear interactions: the trajectories of a self-interacting particle
in a classical or in a quantum background are different.
We finish the Section by proposing a background independent interpretation of the 
propagation in the fuzzy space. We finally conclude with some discussions and perspectives.

\section{Quantum geometry as a fuzzy space}
It has been argued that the quantum dynamics of a scalar field coupled to three dimensional Euclidean gravity 
is ``equivalent" to the dynamics 
of a scalar field (with no gravity at all) evolving in a non-commutative three dimensional space.
This result has originally been illustrated in the context of covariant spin-foam models coupled to massive spinless 
particles \cite{FL}; it was then recovered in the canonical LQG point of view \cite{Noui,Noui1} where the non-commutative
space is constructed such that it admits the Drinfeld double $DSU(2)$ as its isometry algebra. In fact, $DSU(2)$
is a deformation of the (group algebra of the) classical Lie group $ISU(2)$. 

If one assumes that $DSU(2)$ is effectively the isometry algebra of space at the Planck scale, then 
the construction of the quantum geometry at the Planck scale is very similar to constructions of
model spaces in standard classical geometry.
Indeed, in the classical situation, a model space is defined by a coset $G/H$
where $G$ is the isometry (Lie) group and $H$ a subgroup. 
One can easily adpat this construction to the quantum case, the main important difference being 
that the resulting coset is no longer a manifold. It is implicitely
defined by its space of functions, denoted generically $C$ in the sequel, which is endowed with an 
algebra structure. This algebra contains all the geometrical informations of the non-commutative space.

In our specific case, we showed \cite{Noui} that $C$ is the algebra of distributions on the Lie 
group $SU(2)$ endowed with the convolution product. 
This is a result of the Hopf algebra duality which allows to define in
a canonical way an algebraic structure to a (commutative or non-commutative) geometry when its (classical or
quantum) symmetry algebra admits a Hopf algebra structure (in fact we just need a co-product). 
However, this description is rather theoritical and 
it is necessary to know how to explicitely get the geometrical informations out of it. 
This has been done in a companion paper \cite{Noui} and 
this section aims precisely at recalling some of the results obtained in this paper.
First, we show that the quantum geometry described by $C$ is fuzzy in the sense that $C$ is isomorphic to
an algebra of complex matrices; 
then, we exhibit a (non-trivial) link between $C$ and the space $C(\mathbb E^3)$
of functions on the classical space $\mathbb E^3$; we finish with some properties concerning differential calculus 
on $C$, namely we define an integration on $C$ and derivative operators which are necessary to construct
an action for non-commutative fields. Finally, we write an action for a non-commutative scalar field and we
study, as an example, the case where the field is free.

\subsection{Construction of the non-commutative space}
For pedagogical purposes, let us start by presenting briefly how the construction works in the 
classical case before going to the quantum case. In the classical context,
$C$ is the pointwise algebra of functions on the classical Euclidean three-dimensional manifold $\mathbb E^3$
and our problem consists in constructing $C$ starting 
from its isometry group algebra $\mathbb C[ISU(2)]$ where $ISU(2)=SU(2) \ltimes \mathbb R^3$ is 
the Euclidean group. We consider 
the group algebra instead of the group itself to be closer to the quantum case (we will present in the sequel).

The solution is simple. We start by introducing  the space of $SU(2)$-invariant 
linear forms on $\mathbb C[ISU(2)]$ which is, by definition, the space $C(\mathbb R^3)^*$ of distributions
on $\mathbb R^3$. 
The product $\circ$ between two such distributions $f_1$ and $f_2$ is defined from 
the grouplike coproduct $\Delta$ 
on $\mathbb C[ISU(2)]$ using the Hopf algebra duality principle as follows:
\begin{eqnarray}
f_1 \circ f_2 (a)\; = \; (f_1 \otimes f_2)\Delta(a) \;\;\;\; \text{for any $a \in ISU(2)$.}
\end{eqnarray}
Doing so, we obtain, after some trivial calculations, that $\circ$ is the standard convolution product 
in $\mathbb R^3$ and then we have constructed the convolution algebra of distributions on $\mathbb R^3$.
Finally, the algebra $C$ is easily obtained from $C(\mathbb R^3)^*$ performing a  standard Fourier transform 
\cite{Noui}. This closes the classical construction.

\medskip

Let us now present how to adapt the previous construction when $\mathbb C[ISU(2)]$ is replaced by the
quantum double $DSU(2)$. This idea is motivated by the fact that,
in three dimensions, quantum gravity is argued to turn classical isometry group algebras into quantum 
groups \cite{Schroers}. In particular, $\mathbb C[ISU(2)]$ is deformed into $DSU(2)$ when quantizing 
three dimensional Euclidean gravity without cosmological constant, the quantum deformation parameter 
being the Planck length $\ell_P$.
Any element of $DSU(2)$ can be written as $(f\otimes u)$ where $f\in C(SU(2))$ (one can extend the definition
to distributions) are interpreted as ``deformed" translational elements and $u\in \mathbb C[SU(2)]$; when 
$u\in SU(2)$ it is interpreted as a rotational element. 

Following the classical construction, we claim that the space of $SU(2)$-invariant linear forms on $DSU(2)$
is a representation of $C$. This space can be identified with the
convolution algebra $C(SU(2))^*$ of distributions on $SU(2)$: 
its algebra structure has been obtained,
as in the classical case, from the Hopf algebra duality procedure. 
The duality bracket between a distribution $\phi \in C(SU(2))^*$ and a function $f$ in $C(SU(2))$
will be denoted $\langle f,\phi\rangle$ in the sequel. When $\phi$ is a function, the duality bracket
can be given in terms of the (normalized) $SU(2)$ Haar-measure $d\mu$ as follows:
\begin{eqnarray}
\langle f,\phi\rangle \, = \, \int d\mu(u) \, \overline{f(u)} \, \phi(u) \;.
\end{eqnarray} 
Of course, $C$ is a non-commutative algebra which can be interpreted as a 
deformation of the classical algebra $C(\mathbb R^3)^*$ of distributions on the momenta space $\mathbb R^3$. 
It is, in fact, well-known that the momenta space of a particle becomes curved in three dimensional 
Euclidean (quantum) gravity and the standard momenta space is replaced by the Lie group $SU(2)$. 
By construction, $C(SU(2))^*$ provides a representation space of $DSU(2)$ which can be interpreted, in that way,
as a symmetry algebra of $C(SU(2))^*$ whose action will be denoted $\rhd$.
More precisely, translations
elements are functions on $SU(2)$ and acts by multiplication on $C(SU(2))^*$ whereas rotational elements are 
$SU(2)$ elements and act by the adjoint action:
\begin{eqnarray}
\forall \; \phi \in C(SU(2))^* \;\;\; f \rhd \phi = f \phi \;\;\; \text{and} \;\;\; 
u \rhd \phi = \text{Ad}_u \phi \;. 
\end{eqnarray}
The adjoint action is defined by the relation $\langle f,\text{Ad}_u\phi \rangle=
\langle \text{Ad}_{u^{-1}}f,\phi \rangle$ with $\text{Ad}_u f(x)=f(u^{-1}xu)$ for any $u,x$ in $SU(2)$.

\subsection{The fuzzy space formulation}

Thus, we have a clear definition of the deformed space of momenta.
To get the quantum analoguous of the space $C(\mathbb E^3)$ itself, 
we need to introduce a Fourier transform on $C(SU(2))^*$. This is done making use of harmonic analysis 
on the group $SU(2)$:
the Fourier transform of a given $SU(2)$-distribution is the decomposition of that distribution into 
(the whole set or a subset of) 
unitary irreducible representations (UIR) of $SU(2)$. These UIR are labelled by a spin $j$, they are finite dimensional
of dimension $d_j=2j+1$.  The Fourier transform is an algebra morphism which is explicitely defined by:
\begin{eqnarray}
{\cal F}: \;\; C(SU(2))^* & \longrightarrow & \text{Mat}(\mathbb C) \equiv 
\bigoplus_{j=0}^\infty \text{Mat}_{d_j}(\mathbb C)  \\
\phi & \longmapsto & \widehat{\Phi}\equiv{\cal F}[\phi]= \oplus_j {\cal F}[\phi]^j \, = \, \oplus_j (\phi \circ D^j)(e)
\end{eqnarray}
where $\text{Mat}_{d}(\mathbb C)$ is the set of $d$ dimensional complex matrices, $D^j_{mn}$ are
the Wiegner functions and $\circ$ is the convolution product.
When $\phi$ is a function, its Fourier matrix components are obtained performing the following integral 
\begin{eqnarray}
{\cal F}[\phi]^j_{mn} \; \equiv \; \int d\mu(u) \, \phi(u) \, D^j_{mn}(u^{-1}) \,.
\end{eqnarray}
The inverse map ${\cal F}^{-1}:\text{Mat}(\mathbb C)\rightarrow C(SU(2))^*$ associates to any family of matrices
$\widehat{\Phi}=\oplus_j \widehat{\Phi}^j$ a distribution according to the formula:
\begin{eqnarray}
\langle f, {\cal F}^{-1}[\widehat{\Phi}] \rangle \; = \;  \sum_j d_j 
\int d\mu(u) \overline{f(u)} \, \text{tr}(\widehat{\Phi}^j\;D^j(u)) \, 
\equiv \, \int d\mu(u) \overline{f(u)} \,\text{Tr}(\widehat{\Phi}D(u)) 
\end{eqnarray}
for any function $f \in C(SU(2))$. We have introduced the notations $D=\oplus_j D^j$ and 
$\text{Tr}\widehat{\Phi}=\sum_j d_j \text{tr}(\widehat{\Phi}^j)$.
Therefore, it is natural to interpret the algebra 
$\text{Mat}(\mathbb C)$ as a deformation of the classical algebra 
$C(\mathbb E^3)$ and then 
three dimensional Euclidean quantum geometry is fundamentally non-commutative and fuzzy.

\subsection{Relation to $C(\mathbb E^3)$: the non-commutative algebra $C_{\ell_P}(\mathbb E^3)$}
It is not completely trivial to view how the algebra of matrices $\text{Mat}(\mathbb C)$ is a deformation of
the classical algebra of functions on $\mathbb E^3$. 

To make it more concrete, it is necessary to construct a precise link between $C(SU(2))^*$ and
$C(\mathbb R^3)^*$ for the former space is supposed to be a deformation of the later. First,
we remark that it is not possible to find a vector space isomorphism between them because $SU(2)$
and $\mathbb R^3$ are not homeomorphic: in more physical words, there is no way to establish a one
to one mapping between distributions on $SU(2)$ and distributions on $\mathbb R^3$ for $SU(2)$ and
$\mathbb R^3$ have different topologies. Making an explicit link between these two spaces is in fact
quite involved and one construction has been proposed in a companion paper \cite{Noui}. The aim of this 
Section is to recall only the main lines of that construction; more details can be found in 
\cite{Noui}. For pedagogical reasons, we also restrict the space $C(SU(2))^*$ to its subspace  
$C(SU(2))$ and then we are going to present the link between $C(SU(2))$ and $C(\mathbb R^3)$.
\begin{enumerate}
\item First, we need to introduce a parametrization of $SU(2)$: $SU(2)$ is identified with 
$S^3=\{(\vec{y},y_4)\in \mathbb R^4 \vert y^2 + y_4^2=1\}$ and any $u\in SU(2)$ is given by
\begin{eqnarray}
u(\vec{y},y_4)\, = \, y_4-i\vec{y}\cdot \vec{\sigma}
\end{eqnarray}
in the fundamental representation in terms of the Pauli
matrices $\sigma_i$. For later convenience, we cut $SU(2)$ in two parts: the north hemisphere $U_+$ ($y_4>0$) 
and the south hemisphere $U_-$ ($y_4<0$).
\item Then, we construct bijections between the spaces $U_\pm$ and the open ball of $\mathbb R^3$
$B_{\ell_P}=\{\vec{p} \in \mathbb R^3 \vert p<\ell_P^{-1}\}$: to each element $u\in U_\pm$ we 
associate a vector $\vec{P}(u)=\ell_P^{-1}\vec{y}$. 
This bijections implicitely identify $\vec{P}(u)$ with the physical momenta of the theory.
Note that this is a matter of choice: on could have chosen another expression for $\vec P(u)$ and there is no
physical arguments to distinguish one from the other. We made what seems to be, for different reasons, the
more natural and the more convenient choice.
\item As a consequence, any function $\phi\in C(SU(2))$ is associated to a pair of functions 
$\phi_\pm \in C(U_\pm)$, themselves being associated, using the previous bijections, to a pair
of functions $\psi_\pm \in C_{B_{\ell_P}}(\mathbb R^3)$ which are functions on $\mathbb R^3$ with support
on the ball $B_{\ell_P}$. In that way, we construct two mappings 
$\mathfrak a_\pm:C(U_\pm)\rightarrow C_{B_{\ell_P}}(\mathbb R^3)$ such that $\mathfrak a_\pm(\phi_\pm)=\psi_\pm$
are explicitely given by:
\begin{eqnarray}
\psi_\pm(\vec{p}) = \int d\mu(u) \delta^3(\vec p - \vec P(u)) \phi_\pm(u) =
\frac{v_{\ell_P}}{\sqrt{1-{\ell_P^2 p^2}}}
\phi(u({\ell_P\vec{p}},\pm\sqrt{1-{\ell_P^2 p^2}}))
\end{eqnarray}
where $v_{\ell_P}=\ell_P^3/2\pi^2$.
Then we have established a vector space isomorphism $\mathfrak a=\mathfrak a_+ \oplus \mathfrak a_-$
between $C(SU(2))$ and $C_{B_{\ell_P}}(\mathbb R^3)\oplus C_{B_{\ell_P}}(\mathbb R^3)$. We need two functions
on $\mathbb R^3$ to characterize one function of $C(SU(2))$. The mapping $\mathfrak a_\pm$ satisfies the important
following property: the action of the Poincar\'e group $ISU(2)\subset DSU(2)$ on $C_{B_{\ell_P}}(\mathbb R^3)$
induced by the mappings $a_\pm$ is the standard covariant one, namely
\begin{eqnarray}
\xi \rhd \mathfrak a_\pm(\phi_\pm) \; = \; \mathfrak a_\pm(\xi \rhd \phi_\pm) \;\;\;\;\;\;
\forall \, \xi \in ISU(2) \subset DSU(2) \;.
\end{eqnarray}
In the r.h.s. (resp. l.h.s.), $\rhd$ denotes the action of $\xi \in ISU(2)$ (resp. $\xi$ viewed as an element of 
$DSU(2)$) on $C(\mathbb R^3)$ (resp. $C(SU(2))$).
This was in fact the defining property of the mappings $\mathfrak a_\pm$.
\end{enumerate}
Now, we have a precise relation between $C(SU(2))$ and $C(\mathbb R^3)$. Using the standard Fourier transform
$\mathfrak F:C(\mathbb R^3)^*\rightarrow C(\mathbb E^3)$ restricted to $C_{B_{\ell_P}}(\mathbb R^3)$, one obtains
the following mapping:
\begin{eqnarray}
\mathfrak m \, \equiv \, \mathfrak F \circ \mathfrak a \; : \; C(SU(2)) \, \longrightarrow \, C_{\ell_P}(\mathbb E^3)
\end{eqnarray}
where $C_{\ell_P}(\mathbb E^3)$ is defined as the image of $C(SU(2))$ by $\mathfrak m$. 
It will be convenient to introduce the obvious notation $\mathfrak m=\mathfrak m_+ \oplus \mathfrak m_-$.
We have the vector spaces isomorphism 
$C_{\ell_P}(\mathbb E^3) \simeq \widetilde{C}_{B_{\ell_P}}(\mathbb R^3) \oplus \widetilde{C}_{B_{\ell_P}}(\mathbb R^3)$
where $\widetilde{C}_{B_{\ell_P}}(\mathbb R^3)$ is the subspace of functions on $\mathbb E^3$ whose spectra is strictly
contained in the open ball $B_{\ell_P}$ of radius $\ell_P^{-1}$. Elements of $C_{\ell_P}(\mathbb E^3)$ 
are denoted $\Phi_+\oplus \Phi_-$ where $\Phi_\pm(x) \in  \widetilde{C}_{B_{\ell_P}}(\mathbb R^3)$. 
The explicit relation between $C(SU(2))$ and $C_{\ell_P}(\mathbb E^3)$ is
\begin{eqnarray}
\Phi_\pm(x) \, \equiv \, \mathfrak m_\pm(\phi_\pm)(x) \, = \, \int d\mu(u) \phi_\pm(u) \, \exp(iP(u)\cdot x) \;.
\end{eqnarray}
This transform is clearly invertible.
Note that, in \cite{Noui}, 
$C_{\ell_P}(\mathbb E^3)$ is the image of the whole algebra of distributions $C(SU(2))^*$: in that case, 
$C_{\ell_P}(\mathbb E^3)$ is the direct sum of three sub-spaces of $C(\mathbb E^3)$, two of them being isomorphic
to the space of distributions on $\mathbb E^3$ with support on $B_{\ell_P}$, the last one being isomorphic
to the space of distributions on $\mathbb E^3$ with support on $\partial B_{\ell_P}$.

\medskip

It remains to establish the link between $C_{\ell_P}(\mathbb E^3)$ and the space of matrices 
$\text{Mat}(\mathbb C)$. To do so, we make use of the mapping $\cal F$ between $C(SU(2))$ and $\text{Mat}(\mathbb C)$
and the mapping $\mathfrak m$ between the same $C(SU(2))$ and $C_{\ell_P}(\mathbb E^3)$. If we denote by 
$\widehat{\Phi}_\pm$ the images of $\phi_\pm$ by $\cal F$ then we have:
\begin{eqnarray}\label{relPhiMat}
\Phi_\pm(x) \; = \; \text{Tr}(K_\pm^\dagger(x)\widehat{\Phi}_\pm)
\end{eqnarray}
where $K_\pm$ can be interpreted as the components of the element 
$K\equiv K_+\oplus K_- \in \text{Mat}(\mathbb C) \otimes C_{\ell_P}(\mathbb E^3)$  defined by the integral:
\begin{eqnarray}\label{def de K}
K_\pm(x) \, \equiv \, \int_{U_\pm} d\mu(u) \, D(u) \, \exp(-iP(u)\cdot x) \;.
\end{eqnarray}
\medskip
The relation (\ref{relPhiMat}) is invertible. One can interpret the functions $\Phi_\pm(x)$ as a kind of continuation
to the whole Euclidean space of the discrete functions $\widehat{\Phi}^j_{\pm mn}$ which are a priori defined only on
a infinite but numerable set of points. Given $x \in \mathbb E^3$, each matrix element $\widehat{\Phi}^j_{\pm mn}$
contributes to the definition of $\Phi_\pm(x)$ with a complex weight $\overline{K^j_{\pm nm}(x)}$.

\medskip

For the moment, we have only described the vector space structure of $C_{\ell_P}(\mathbb E^3)$.
However, this space inherits a non-commutative algebra structure when we ask the mapping $\mathfrak m$
to be an algebra morphism. The product between two elements $\Phi_1$ and $\Phi_2$ in $C_{\ell_P}(\mathbb E^3)$
is denoted $\Phi_1\star\Phi_2$ and is induced from the convolution product $\circ$ on $C(SU(2))$ as follows:
\begin{eqnarray}
\Phi_1\star\Phi_2 \; = \; \mathfrak m(\mathfrak m^{-1}(\Phi_1) \circ \mathfrak m^{-1}(\Phi_2))\;.
\end{eqnarray}
The $\star$-product is a deformation of the classical pointwise product. A very similar $\star$-product
has been introduced in \cite{FL} in the context of Spin-Foam models; the main difference being that their algebra
consists in only one copy of $C_{B_{\ell_P}}(\mathbb E^3)$ and then appears to be not clearly related 
to $C(SU(2))$. 

In order to make the $\star$-product more intuitive, 
it might be useful to consider some examples of products of functions. 
The more interesting functions to consider first are surely the plane waves.
Unfortunately, plane waves are not elements of $C(SU(2))$ but are pure distributions and then,
their studies goes beyond what we recalled in this paper. Nevertheless, we will see that it is 
possible to extend the previously presented results to the case of the plane waves with some assumptions.
Plane waves are defined as eigenstates of the generators $P_a$ and then, as we have already underlined,
a plane wave is represented by the distribution
$\delta_u$ with eigenvalue $P_a(u)$ which is interpreted as the momentum of the plane wave. 
Plane waves are clearly degenerated as $P_a(u)$ is not invertible in $SU(2)$: this result illustrates 
the fact that we need two functions $\Phi_+\oplus \Phi_- \in C_{\ell_P}(\mathbb E^3)$ to characterize one function 
$\phi \in C(SU(2))$.
The representations
of the plane wave in the matrix space $\text{Mat}(\mathbb C)$ and in the continuous space
${C}_{\ell_P}(\mathbb E^3)$ are respectively given by:
\begin{eqnarray}
{\cal F}(\delta_u)^j \; = \; D^j(u){}^{-1} \;\;\;\; \text{and}  \;\;\;\;
{\mathfrak m}(\delta_u)(x) \; \equiv \; w_u(x)
\end{eqnarray}
where $w_u(x)=\exp(iP_a(u)x^a) \oplus 0$ if $u\in U_+$ and $w_u(x)=0 \oplus \exp(iP_a(u)x^a)$ if $u\in U_-$.
The framework we have described do not include the case $u\in \partial U_+=\partial U_-$ which is nonetheless
completely considered in \cite{Noui}.
The $\star$-product between two plane waves reads:
\begin{eqnarray}
w_u \; \star \; w_v \; = \; w_{uv}
\end{eqnarray}
if $u$, $v$ and $uv$ belongs to $U_+$ or $U_-$. This product can be trivially extended
to the cases where the group elements belong to the boundary $\partial U_+=\partial U_-$.
As a result, one interprets $P_a(u) \boxplus P_a(v) \equiv P_a(uv)$ as the deformed addition rule of 
momenta in the non-commutative space.

Other interesting examples to consider are the coordinate functions. 
They are easily defined using the plane waves and their definition in the $C(SU(2))^*$ and
$\text{Mat}(\mathbb C)$ representations are:
\begin{eqnarray}
\chi_a =2i\ell_P \xi_a \delta_e \in C(SU(2))^* \;\;\;\;\;\;\;\;
\widehat{x}_a = 2\ell_P D(J_a) \in \text{Mat}(\mathbb C)
\end{eqnarray}
where $\xi_a$ is the $SU(2)$ left-invariant vector field and $J_a$ the generators of the $\mathfrak{su}(2)$
Lie algebra satisfying $[J_a,J_b]=2i\epsilon_{ab}{}^c J_c$. In the $C_{\ell_P}(\mathbb E^3)$ representation,
the coordinates are given by $X_a\equiv(x_a\oplus 0)$; only the first component is non-trivial.
It becomes straightforward to show that the coordinates satisfy the relation
\begin{eqnarray}
[X_a,X_b]_\star \; \equiv \; X_a  \star X_b  - 
X_b \star X_a \; =\; i\ell_P \epsilon_{ab}{}^c X_c 
\end{eqnarray}
and therefore do not commute as expected.

\medskip 

We end this Section by a quick summary of the different representations of the (suitable sub-algebra of the)
algebra $C$:
giving a function $\phi \in C(SU(2))$ is equivalent to give either a couple of functions
$\psi_\pm =\mathfrak a_\pm(\phi)$ which belong to $ C(\mathbb R^3)$; or a couple of functions 
$\Phi_\pm=\mathfrak m_\pm(\phi) \in C(\mathbb E^3)$; or a matrix 
$\widehat{\Phi}={\cal F}(\phi) \in \text{Mat}(\mathbb C)$ or finally a couple of matrices 
$\widehat{\Phi}_\pm={\cal F}(\phi_\pm) \in \text{Mat}(\mathbb C)$. 

\subsection{An integral on the non-commutative algebra}
An important property is that the non-commutative space admits an invariant measure 
$h:C \rightarrow \mathbb C$. To be more precise, $h$ is well defined  
on the restriction of $C \simeq C(SU(2))^*$ to $C(SU(2))$.
The invariance is defined with
respect to  the symmetry action of the Hopf algebra $DSU(2)$. Let us give the expression of
this invariant measure in the different formulations of the non-commutative space:
\begin{eqnarray}\label{measure}
h(\phi) \; = \; \phi(e) \; = \; \text{Tr}(\widehat{\Phi}) \; = \; 
 \int \frac{d^3x}{(2\pi)^3v_{\ell_P}} \; \Phi_+(x) 
\end{eqnarray}
where $\phi \in C(SU(2))$, $\widehat{\Phi}={\cal F}[\phi]$ and $\Phi_+(x)={\mathfrak m}_+[\phi](x)$.
Note that $\int d^3x$ is the standard Lebesgue measure on the classical manifold $\mathbb E^3$. 
This measure can be extended to distributions which are well-defined at the origin $e$ in the sense that they 
behave like regular functions at the vicinity of $e$.

Sometimes, such a measure is called a trace. It allows to define a norm on the algebra $C$ from
the hermitian bilinear form 
\begin{eqnarray}
\langle \phi_1,\phi_2\rangle \; \equiv \; h({\phi}^{\flat}_1\phi_2) \; = \; 
\int d\mu(u) \, \overline{\phi_1(u)}\phi_2(u)
\end{eqnarray}
where $\phi^{\flat}(u)=\overline{\phi(u^{-1})}$. As we will see below, such a trace is necessary to define
an action for a field living on the non-commutative space.

\subsection{Derivative operators}
Derivative operators $\partial_\xi$ can be deduced from the action of infinitesimal translations: 
given a vector $\xi \in \mathbb E^3$, we have $\partial_\xi=\xi^a\partial_a$
where $\partial_a=iP_a$ is the translation operator we have introduced in the previous section.
When acting on the $C(SU(2))$ representation, $\partial_\xi$ is the multiplication
by the function $i\xi^aP_a$; it is the standard derivative
when acting on the continous ${C}_{\ell_P}(\mathbb E^3)$ representation (using the mapping $\mathfrak m$); 
finally it is a finite difference operator
when acting on the fuzzy space representation $\text{Mat}(\mathbb C)$ (using the Fourier transform $\cal F$).
After some calculations, one shows that its expression in the matrix representation is then given by the 
following:
\begin{eqnarray}
(\partial_a \widehat{\Phi})^j   & = &  
\text{tr}_{j-1/2}\left[(\widehat{\Phi}^{j-1/2} \otimes \mathbb I) \cdot C_a(j-1/2,j)\right] \nonumber \\
 && +  \text{tr}_{j+1/2}\left[(\widehat{\Phi}^{j+1/2}\otimes \mathbb I) \cdot C_a(j+1/2,j)\right]\nonumber
\end{eqnarray}
where we have introduced the operator 
$C_a(j,k) \in \text{Mat}_{d_j}(\mathbb C) \otimes \text{Mat}_{d_k}(\mathbb C)$:
\begin{eqnarray} 
C_a(j,k) \; \equiv \; -\frac{1}{\ell_P} \int d\mu(u) 
\; \text{tr}[D^{1/2}(J_au)] \; \overline{D^j(u)} \otimes  D^k(u) \;.
\end{eqnarray}
The notation $\text{tr}_j$ means that we perform a trace in 
the space of dimension $d_j$.
The matrix coefficients of the operator $D_a(j,k)$ can be explicitely computed in terms of $SU(2)$ Clebsh-Gordan coefficients and we finally get the following operator
\begin{eqnarray}\label{derivative}
(\partial_a \widehat{\Phi})^j_{st} \; = \; -\frac{1}{\ell_P d_j} D^{1/2}_{pq}(J_a) && 
\!\!\!\!\!( \sqrt{(j+1+2qs)(j+1+2tp)} \; \widehat{\Phi}^{j+1/2}_{q+s \, p+t} \nonumber\\
&& + (-1)^{q-p}\sqrt{(j-2qs)(j-2pt)} \; \widehat{\Phi}^{j-1/2}_{q+s \, p+t} ).
\end{eqnarray} 
Details of the calculation can be found in the appendix of the companion paper \cite{Noui}.
The interpretation of the formula (\ref{derivative}) is clear. Note however an important point: 
the formula (\ref{derivative}) defines a second order operator in the sense that it involves
$\widehat{\Phi}^{j-1/2}$ and $\widehat{\Phi}^{j+1/2}$ that are not nearest matrices but second nearest matrices.

The derivative operator is obviously necessary to define a dynamics in the non-commutative fuzzy space.
The ambiguity in the definition of $P_a$ implies immediately an ambiguity in the dynamics. 
For instance, the fact that $C_a(j,k)$
relates matrices $\widehat{\Phi}^j$ with $\widehat{\Phi}^{j\pm 1/2}$ only is a consequence
 of the choice of $P_a$ which is in fact a function
whose non-vanishing Fourier modes are the matrix elements of a dimension 2 matrix: 
indeed, $P_a(u)=\ell_P^{-1}\text{tr}_{1/2}(J_au)$. 
Another choice would lead to a different dynamics and then there is ambiguity. 
Such ambiguities exists as well in full LQG.

\subsection{Free field: solutions and properties}
Now, we have all the ingredients to study dynamics on the quantum space.
Due to the fuzzyness of space, equations of motion will be discrete and therefore, there is in general no equivalence
between Lagrangian and Hamiltonian dynamics. Here, we choose to work in the Euler-Lagrange point of view, i.e.
the dynamics is governed by an action of the type:
\begin{eqnarray}\label{action}
S_\star[\Phi,J] \; = \; \frac{1}{2}
\int \frac{d^3x}{(2\pi)^3 v_{\ell_P}} \; 
\left(\partial_\mu \Phi \star \partial_\mu \Phi \; + \; V(\Phi,J)\right)_+(x)
\end{eqnarray}
where $V$ is the potential that depends on the field $\Phi$ and eventually on some exterior fields $J$.
The action has been written in the ${C}_{\ell_P}(\mathbb E^3)$ formulation to mimic easily the classical situation.
However, one has to be aware that $\Phi$ comes from an element $\phi \in C(SU(2))$ in the sense that 
$\Phi(x)={\mathfrak m}(\phi)(x)$ and therefore cannot be any classical function on $\mathbb E^3$; in particular,
it has a bounded spectrum. The integral we use to define the action is the measure introduced in previous sections
(\ref{measure}). 

Finding the equations of motions reduces obviously in extremizing the previous action, but with the constraint 
that $\Phi$ belongs to ${C}_{\ell_P}(\mathbb E^3)$: in particular, $\Phi$ (as well as the exterior field) 
admits two independent components 
$\Phi_\pm$ which are classical functions on $\mathbb E^3$ whose spectra are bounded. The action 
(\ref{action}) couples generically these two components.
Even when one of the two fields vanishes, for instance $\Phi_-=0$, 
it happens in general that the extrema of the functional
$S[\Phi]$ differ from the ones that we obtain for a classical field $\Phi$ whose action would be formally the same
functional but defined with the pointwise product instead of the $\star$ product. This makes the classical solutions
in the deformed and undeformed cases different in general.
Let us precise this point. When the field is free in the sense that $V$ is quadratic (with a mass term), deformed solutions
are the same as classical ones. However, solutions are very different when the dynamics
is non-linear and the differences are physically important. It is the purpose of this paper to illustrate this fact
in some simple examples.

\medskip

First, let us consider the case of a free field: we assume that $V(\Phi)=\mu^2\Phi\star \Phi$
where $\mu$ is a positive parameter. 
Equations of motion are obtained by extremizing the action with the constraints that $\Phi={\mathfrak m(\phi)}$,
$\phi$ being in $C(SU(2))$. These equations are best written in the fuzzy space formulation
and one gets as expected the following set of finite difference equations:
\begin{eqnarray}
{\Delta \widehat{\Phi}}^j\; + \; \mu^2 \widehat{\Phi}^j \; = \; 0 \;\;\; \text{for all spin $j$}.
\end{eqnarray}
Due to the quite complicated expression of the derivative operator, it appears more convenient
to solve this set of equations in the $C(SU(2))$ representation. 
Indeed, these equations are equivalent to the fact that $\phi ={\cal F}^{-1}[\Phi]$ 
has a support in the conjugacy classes $\theta \in [0,2\pi[$
such that $\sin^2(\theta/2)=\ell_P^2\mu^2$. Thus, a solution exists only if $\mu \leq \ell_P^{-1}$, 
in which case we write $\mu=\ell_P^{-1}\sin (m/2)$  with $0<m<\pi$.
Then the solutions of the previous system are given by $\widehat{\Phi}=\widehat{\Phi}_++\widehat{\Phi}_-$ with: 
\begin{eqnarray}
\widehat{\Phi}_\pm^j \; = \; \int d\mu(u) \; \mathbb I^\pm_m(u) \left( \alpha(u) D^j(u) \; + \; 
\beta(u) D^j(u)^\dagger \right) 
\end{eqnarray}
where $\alpha$ and $\beta$ are $SU(2)$ complex valued functions;
the notation $\mathbb I^\pm_m$ holds for the caracteristic functions on the conjugacy class $\theta=m$ (for the $+$ sign)
and $\theta=2\pi-m$ (for the $-$ sign). These functions are normalized
to one according to the relation $\int d\mu(u) \mathbb I^\pm_m(u)=1$.
If the fields $\Phi_\pm(x)$ are supposed to be real, 
the matrices $\widehat{\Phi}^j$ are hermitian, and then $\alpha$ and $\beta$ are complex conjugate functions.
As a result, we obtain the general solution for the non-commutative free field written in the fuzzy space 
representation. 

Using the mapping $\mathfrak m$, one can reformulate this solution in terms of functions on $\mathbb E^3$. 
The  components of $\Phi$ are given by:
\begin{eqnarray}
\Phi_\pm(x) \; = \; 
\frac{\ell_P^2}{16\pi}\frac{\sin^2\frac{m}{2}}{\cos\frac{m}{2}}\int_{B_{\ell_P}} \!\!\! d^3p \,
\delta(p-\mu) \; \left( {\alpha}_\pm(p) 
e^{{i}p\cdot x} 
\; + \; {\beta}_\pm(p) e^{-{i}p\cdot x}\right)
\end{eqnarray}
where $B_{\ell_P}$ is the Planck ball, ${\alpha}_\pm(p)=\alpha(u(p))$ where $u(p)$ is the inverse of $p(u)$ when 
$u$ is restricted to the sets $U_\pm$; a similar definition holds for $\beta_\pm$.
We recover the 
usual solution for classical free scalar fields with  the fact that the mass has an upper limit given by $\ell_P^{-1}$. 
Therefore, the Planck mass appears to be a natural UV cut-off.
This result can a priori be extended to any free (quadratic) field theory, like Dirac or Maxwell theory for instance.
We hope to study these important examples in future works.

\section{Particles evolving in the fuzzy space}
Important discrepencies between classical and fuzzy dynamics appear when one considers non-linear interactions.
In the case we study the dynamics of a sole field $\phi$, one has to introduce self-interactions. 
However, even in the standard classical commutative space $\mathbb E^3$, 
classical solutions of self-interacting field cannot be written in a closed form in general; and then one cannot expect 
to find explicit solutions for the self-interacting field evolving in the fuzzy background. 
Face with such technical difficulties (that we postpone for future investigations), 
we will consider simpler models. We will perform symmetry reductions in order 
that the field $\phi$ depends only on one coordinate out of the three.
We will interpret this model as describing one particle evolving in (Euclidean) fuzzy space-time.

\subsection{Reduction to one dimension}
Let us define the algebra $C^{1D}$ of symmetry reduced fields and its different 
representations: the group algebra, the matrix and the continuous formulations. Using a trivial analogy
with the classical case, $C^{1D}$ is defined as the kernel of the operators 
$P_1$ and $P_2$ in the convolution algebra $C(SU(2))^*$ where $P_a$ are the momentum coordinates:
\begin{eqnarray}\label{def de C1D}
C^{1D} \; \simeq \; \{ \phi \in C(SU(2))^* \; \vert \; \phi=\varphi(P_1)\delta(P_2)\delta(P_3) \} \,.
\end{eqnarray}
As a result, $C^{1D}$ can be identified to the set $C(U(1))^*$ of $U(1)$ distributions. 
This set inherits an algebra structure from the product on the full algebra $C$:
it is the $U(1)$ convolution product.
Note that, the algebra becomes commutative but, as we will see in the sequel, the product is still non-trivial
and exhibits interesting properties compared to the classical one. In the sequel, we identify $\phi$ of 
$C^{1D}$ with the $U(1)$ distributions $\varphi$ (\ref{def de C1D}) and we choose a parametrization such that 
$\varphi$ is a function of $\theta \in [0,2\pi]$.
The algebra $C^{1D}$ admits two other formulations: the matrix one obtained from the 
induced Fourier transform and
the continuous one obtained from the induced map $\mathfrak m$. 

Let us first consider the matrix representation.
A priori, any element $\varphi \in C^{1D}$  admits as a Fourier transform an infinite set of matrices.
This set is in fact highly degenerate due to the symmetry reduction and reduces to only one infinite dimensional
diagonal matrix $\widehat{\Phi} \in \text{Diag}_\infty(\mathbb C)$. The relation between the diagonal matrix elements
$\widehat{\Phi}^a_a$ and the associated distribution $\varphi$ is given by:
\begin{eqnarray}
{\cal F}^{1D} \; : \; C(U(1))^* \; \longrightarrow \; \text{Diag}_\infty(\mathbb C) \;\;,\;\;\;\;
\varphi \; \longmapsto \; \widehat{\Phi} \;\; \text{with} 
\;\;\widehat{\Phi}^a_a \; \equiv \; \varphi_a \; = \; \langle \varphi , e^{ia\theta}\rangle \;
\end{eqnarray}
where $\langle,\rangle$ is the duality bracket between $U(1)$ distributions and $U(1)$ functions.
This identity reduces  to the more concrete following relation when $\varphi$ is supposed to be a 
function:
\begin{eqnarray}
\varphi_a \; = \; \frac{1}{2\pi}\int_{0}^{2\pi}\!\!{d\theta} \; {\varphi}(\theta) e^{ia\theta} \;.
\end{eqnarray}
Thus, the non-commutative Fourier transform reduces to the simple Fourier modes decomposition of
a periodic one-dimensional function.
Indeed, we have $\text{Diag}_\infty(\mathbb C) \simeq \mathbb Z \otimes \mathbb C$
which is the Fourier space of $U(1)$ distributions.
The algebra structure of $\text{Diag}_\infty(\mathbb C)$ is induced from the convolution product $\circ$
and is simply given by the commutative discrete pointwise product:
\begin{eqnarray}
\forall \; \varphi,\varphi' \in C(U(1))^* \;\;\;\;
(\varphi \circ \varphi')_a \; = \; \varphi_a \; \varphi'_a \;.
\end{eqnarray}

\medskip

Let us now construct the mapping between the convolution algebra $C(U(1))$ and 
the algebra ${C}_{\ell_P}(\mathbb E^1)$ which has to be understood for the moment as 
the one-dimensional analoguous of ${C}_{\ell_P}(\mathbb E^3)$. We proceed in the same way as
in the full theory: 
\begin{enumerate}
\item first, we cut $U(1)\equiv[0,2\pi]$ in two parts, $U_+\equiv]-\frac{\pi}{2},\frac{\pi}{2}[$ and 
$U_-\equiv]\frac{\pi}{2},\frac{3\pi}{2}[$ where the symbol $\equiv$ means equal modulo $2\pi$; 
\item then, we construct two bijections between $U_\pm$ and 
$B_{\ell_P}^{1D}\equiv]-\ell_P^{-1};\ell_P^{-1}[$ by assigning to each $\theta \in U_\pm$ a momentum 
$P(\theta)=\ell_P^{-1}\sin\theta$; 
\item the third step consists in associating to any function $\varphi \in C(U(1))$
a pair of functions $\varphi_\pm \in C(U_\pm)$, and a pair of functions $\psi_\pm \in C(\mathbb R)$ induced by 
the previous bijections as follows
\begin{eqnarray}
\mathfrak a^{1D}_\pm(\phi_\pm)(p)\equiv 
\psi_{\pm}(p) & = & \int\frac{d\theta}{2\pi} \, \delta(p-\ell_P^{-1}\sin\theta) \, \varphi_\pm(\theta) \nonumber \\
& = & \frac{\ell_P}{4\pi}\frac{1}{\sqrt{1-\ell_P^2p^2}} \varphi_\pm(\theta(p))
\end{eqnarray}
where $\theta(p)$ is the inverse of $p(\theta)=\ell_P^{-1}\sin\theta$ in each open $U_\pm$;
\item finally, we make use of the standard one dimensional Fourier transform $\mathfrak F^{1D}$ to 
construct the mapping $\mathfrak m^{1D}=\mathfrak m_+^{1D} \oplus  \mathfrak m_-^{1D}:C(U(1))\rightarrow 
C_{\ell_P}(\mathbb E^1)$ where the components $\mathfrak m_\pm^{1D} = {\mathfrak F}^{1D}\circ \mathfrak a_\pm^{1D}$ 
are given by:
\begin{eqnarray}
\mathfrak m_\pm^{1D}(\varphi_\pm)(t)  \equiv  
\Phi_\pm(t) \, = \, \int_0^{2\pi} \frac{d\theta}{2\pi} \, \varphi_\pm(\theta) \, \exp(iP(\theta)t)\,.
\end{eqnarray}
The space $C_{\ell_P}(\mathbb E^1)$ is the image of $C(U(1))$ by $\mathfrak m$ and therefore
is defined by $\widetilde{C}(U_+) \oplus \widetilde{C}(U_-)$ where $\widetilde{C}(U_\pm)$ are the image
by $\mathfrak F^{1D}$ of ${C}(U_\pm)$. As in the full theory, this construction can be extended to the algebra
$C(U(1))^*$ of distributions.
\end{enumerate}

It remains to construct the link between the discrete and the continuous representations of $C^{1D}$.
To do so, we compose the Fourier transform with the map $\mathfrak m^{1D}$, and we obtain the reduced version of 
the formula
(\ref{relPhiMat}) linking $\Phi_\pm(t)$ with $\varphi_a$:
\begin{eqnarray}\label{relreduced}
\Phi_\pm(t) \; = \; \sum_a \varphi_a \, K^a_\pm(t) 
\end{eqnarray}
where the functions $K_\pm^a(t)$ are defined by the integrals
\begin{eqnarray}
K^a_\pm(t)  \equiv  \int_{U_\pm}\frac{d\theta}{2\pi} e^{-ia\theta+iP(\theta)t} 
 =  (\pm 1)^a \int_0^{\frac{\pi}{2}} \frac{d\theta}{\pi} \, \cos(a\theta \mp \frac{t}{\ell_P}\sin \theta) \;.
\end{eqnarray}
As in the general case, the relation (\ref{relreduced}) is invertible. The integral defining $K_\pm$
is a simplified version of the general formula (\ref{def de K}) and one can viewed these functions as the
components of the element $K=K_+\oplus K_- \in \text{Diag}_\infty(\mathbb C)\otimes C_{\ell_P}(\mathbb E^1)$.
Furthermore, $K^a=K^a_+\oplus K^a_-$ is the image by $\mathfrak m^{1D}$ of the (discrete) plane waves
$\exp(-ia\theta)$. As a last remark, let us underline that $K_+$ and $K_-$ are closely related
by the property $K_-^a(-t)=(-1)^aK_+(t)$. This implies that the functions $\Phi_\pm$ are also closely related:
if we assume for instance that $\varphi_{2n+1}=0$ for any $n \in \mathbb Z$ then $\Phi_-(-t)=\Phi_+(t)$;
if we assume on the contrary that $\varphi_{2n}=0$ for any $n \in \mathbb Z$ then $\Phi_-(-t)=-\Phi_+(t)$.
Such a property will have physical consequences as we will see in the sequel.

Let us give some physical interpretation of the formula (\ref{relreduced}). One can view it as a way to extend
$\varphi_a$, considered as a function on $\mathbb Z$, into the whole real line $\mathbb R$. In that sense,
this formula is a link between the discrete quantum description of a field and a continous classical description.
One sees that any microscopic time $a$ contributes (positively or negatively) to the definition of a macroscopic 
time $t$
with an amplitude precisely given by $K^a_\pm(t)$. At the classical limit 
$\ell_P \rightarrow 0$, $K^a_\pm(t)$ are maximal for values of the time $t=\pm\ell_P a$.
In other words, the more 
the microscopic time $a\ell_P$ is close to the macroscopic time $t$, the more the amplitude $K^a_\pm(t)$
is important. 

Concerning the reduced $\star$-product, it is completely determined by the algebra of the functions $K^a$
viewed as elements of $C_{\ell_P}(\mathbb E^1)$ and a straightforward calculation leads to the following product
between $K^a$ type functions:
\begin{eqnarray}
K^a \star K^b \; \equiv \; \mathfrak m^{1D}(\exp(-ia\theta) \circ \exp(-ib\theta)) \,  = \, \delta^{ab} K_a \,.
\end{eqnarray}
This result clearly illustrates the non-locality of the $\star$-product.

Before going to the dynamics, let us give the expression of the derivative operator $\partial_t$. 
As for the general case, $\partial_t$ is a finite difference operator whose action on 
$\text{Diag}_\infty(\mathbb C)$ is given, as expected, by the following formula:
\begin{eqnarray}
(\partial_t \varphi)_a \; = \; \frac{1}{2\ell_P}\left(\varphi_{a+1} - \varphi_{a-1} \right) \;.
\end{eqnarray}
This expression is highly simplified compared to the more general one introduced in the previous section.
However, we still have the property that $\partial_t$ is in fact a second order operator for it relates $a+1$
and $a-1$. A important consequence would be that the  dynamics (of the free field) will decouple the odd components
$\varphi_{2n}$ and the even components $\varphi_{2n+1}$ of the discrete field. Then, we will have two independent 
dynamics which could be interpreted as two independent particles evolving in the fuzzy space.
In particular, one could associated the continuous fields $\Phi(t)^{odd}$ and $\Phi(t)^{even}$ respectively 
associated to the families $(\varphi_{2n})$ and $(\varphi_{2n+1})$. It is clear that $\Phi(t)^{odd}$ and
$\Phi(t)^{even}$ are completely independent one to the other and, 
using the basic properties of $K_\pm$, we find that the $\pm$
components  of each field are related by:
\begin{eqnarray}
\Phi_-^{odd}(-t) = \Phi_+^{odd}(t) \;\;\;\text{and}\;\;\;
\Phi_-^{even}(-t) = -\Phi_+^{even}(t)\;.
\end{eqnarray}
Thus, $\Phi_+$ and $\Phi_-$ fundamentaly describe two ``miror" particles.

\subsection{Dynamics of a particle: linear vs. non linear}
Now, we have all the ingredients to study the behavior of the one dimensional field $\varphi$. 
When written in the continuous representation, its dynamics is governed by an action of type 
(\ref{action}) but one-dimensional only, with no external field $J$
and the potential is supposed to be monomial, i.e. of the form 
$V(\Phi)=\varepsilon/(\alpha+1) \Phi^{\star (\alpha+1)}$
with $\alpha+1$ a non-null integer. 
The equations of motions are given by:
\begin{eqnarray}
\Delta \Phi \, + \, \varepsilon  \Phi^{\star \alpha} \; = \; 0 
\end{eqnarray}
where $\Delta=\partial_t^2$.
Due to the form of the $\star$-product, these equations are generically $(\alpha>1)$ highly non-local 
and mix the two components $\Phi_\pm$ of the field. In fact, they are best written in the fuzzy space
representation where they reduce to the following finite difference equation:
\begin{eqnarray}\label{dynamics_particle}
\frac{\varphi_{a+2} \; - \; 2\varphi_a\; + \; \varphi_{a-2}}{4\ell_P^2} \; = \;
-\varepsilon \varphi_a^{\alpha} \;.
\end{eqnarray}
As it was previously emphasized, 
we  note that these equations do not couple  odd and even integers $a$. For simplicity purposes,
we will consider only even spins, i.e. we assume that $\varphi_{2n+1}=0$ for all integer $n$.

To warm up, let us start with a simple example: the case where the potential is those of a harmonic oscillator, 
i.e. $\alpha=1$ and $\varepsilon=\Omega^2$. 
In that case, the  system admits a simple exact solution given by:
\begin{eqnarray}
\varphi_a \; = \; a_+ \exp(i\omega_0 \ell_Pa) \; + \; a_- \exp(-i\omega_0 \ell_Pa)
\end{eqnarray} 
where $\omega_0$ satisfies the defining equation $\Omega^2\ell_P^2 =\sin^2(\omega_0\ell_P)$ 
together with (the restriction that) $\omega_0\ell_P\in[0,\pi/2]$ and then we have 
implicitely assumed that $\Omega\ell_P\leq 1$, which means that the period $\Omega^{-1}$ cannot be smaller than 
the Planck time. Otherwise, there is no oscillations and the amplitude of the motion decreases exponentially.
Using the formula (\ref{relreduced}), one can extend this solution to the whole real line and one shows that
the component $\Phi_\pm$ are explicitely given by:
\begin{eqnarray}\label{Phi+part}
\Phi_\pm(t) \; = \; a_+ \exp(\pm iP(\omega_0 \ell_P)t) \; + \; a_- \exp(\mp iP(\omega_0 \ell_P)t)
\end{eqnarray}
where $P(\omega_0\ell_P)=\Omega$. It is interesting to note that the two components are simply related
by $\Phi_+(t)=\Phi_-(-t)$: thus, $\Phi_+$ and $\Phi_-$ have the same physical content; we will give an
interpretation of that property in the sequel.
As expected, the solution for $\Phi_\pm$ is the same as the standard classical one 
where the period of the oscillations is bounded.
Nonetheless, the periods for the discrete field and the continuous field are
different: one can interpret $\Omega$ as a renormalization of $\omega_0$ due to gravitational effects. 

\medskip

This clearly shows that dynamics of a one-dimensional free field in the fuzzy space
is very similar to those in a classical space. 
When the dynamics is non-linear, solutions are no longer the same and this section is devoted to
illustrate this point. 

For that purpose, we consider the dynamics (\ref{dynamics_particle}) with $\alpha \geq 2$ and 
we look for perturbative solutions in the parameter $\varepsilon$. The corresponding classical solution $\Phi_c$ 
reads at the first order
\begin{eqnarray}\label{classical_solution}
\Phi_c(t) \; = \; vt \; - \; \varepsilon \frac{v^\alpha \; t^{\alpha +2}}{(\alpha+1)(\alpha+2)} \; + \; 
{\cal O}(\varepsilon^2)  
\end{eqnarray}
where we assume for simplicity that $\Phi_c(0)=0$ and $\Phi_c'(0)=v$.

The perturbative expansion of the fuzzy solution is obtained using the same techniques. We look for solutions of
the type $\varphi_a=\lambda a+\varepsilon \eta_a$ where $a=2k$ by asumption,
$\lambda$ is a real number and $\eta$ must satisfy the following relation:
\begin{eqnarray*}
\eta_{2k}  -  \eta_{2k-2} & = & -\frac{\ell_P^2}{4} \lambda^\alpha \sum_{n=1}^{k-1} (2n)^\alpha \; = \; 
 -\frac{\ell_P^2}{4} (2\lambda)^\alpha [\frac{(k-1)^{\alpha +1}}{\alpha +1} + \frac{(k-1)^\alpha}{2} \\
 & & + \frac{\alpha(k-1)^{\alpha -1}}{12} - \frac{\alpha(\alpha-1)(\alpha-2)}{720}(k-1)^{\alpha-3}\\
& & + \frac{\alpha(\alpha -1)(\alpha-2)(\alpha-3)(\alpha-4)}{30240}(k-1)^{\alpha-5} + \cdots]
\end{eqnarray*}
The solution is in general complicated.
To be explicit, we will consider the case $\alpha=2$. The formula simplifies a lot,
and after some straightforward calculations, one shows that:
\begin{eqnarray}
\eta_{2k} \; = \; -\frac{\ell_P^2\lambda^2}{12}k^2(k-1)(k+1) \; = \;  
-\frac{\ell_P^2\lambda^2}{12} (k^4-k^2)\,.
\end{eqnarray}
In order to compute the $C_{\ell_P}(\mathbb E^1)$ representation of this solution,
one uses the following relations for any integer $n$
\begin{eqnarray}
S_\pm^{(n)}(t) \; \equiv \; 
\sum_{k=-\infty}^{+\infty} k^n K_\pm^{2k}(t) \; = \; \frac{1}{2(2i)^n} \frac{d^n}{d\theta^n}\exp(iP(\theta)t)
\vert_{\frac{1\mp 1}{2}\pi} \;.
\end{eqnarray}
Applying this formula for $n=1,2$ and $4$
\begin{eqnarray} \label{identity for S}
S_\pm^{(1)}(t)=\pm \ell_P^{-1}t ,\;\;\;
S_\pm^{(2)}(t)=2\ell_P^{-2}t^2 ,\;\;\; 
S_\pm^{(4)}(t)= 2(4\ell_P^{-4}t^4+\ell_P^{-2}t^2)
\end{eqnarray}
one shows, after some simple calculations, that
that $\Phi_+$ and $\Phi_-$  are simply related by $\Phi_+(t)=\Phi_-(-t)$ and $\Phi_+$ is given by:
\begin{eqnarray}
\Phi_+(t) \, = \, 2\lambda \ell_P^{-1}t \, - \, \varepsilon \frac{\ell_P^2 \lambda^2}{6}
(2\ell_P^{-4}t^4 + \ell_P^{-2}t^2) \; + \; {\cal O}(\varepsilon^2)\,.
\end{eqnarray}
To compare it with the classical solution $\Phi_c$ computed above
(\ref{classical_solution}), we impose the same initial conditions which leads to 
$\lambda=v\ell_P/2$ and then the solution reads:
\begin{eqnarray}
\Phi_+(t) \; = \; vt - \varepsilon \frac{v^2t^4}{12} \, - \, \varepsilon \frac{\ell_P^2 v^2 t^2}{24} \, + \, {\cal O}(\varepsilon^2) \;.
\end{eqnarray}
Let us interpret the solution.
First, let us underline once again that $\Phi_+$ and $\Phi_-$ are related by $\Phi_+(t)=\Phi_-(-t)$: 
thus, it seems that $\Phi_-$ corresponds to a particle
evolving backwards compared to $\Phi_+$. In that sense, the couple $\Phi_\pm$ behaves like a particle 
and a "miror" particle: the presence of the miror particle is due to quantum gravity effects.
Second, we remark that the solution for $\Phi_+$  differs from its classical counterpart at least 
order by order in the parameter $\varepsilon$.
At the no-gravity limit $\ell_P \rightarrow 0$, $\Phi_+$ tends to the classical solution
(\ref{classical_solution}). 
Therefore, we can interpret these discrepencies as an illustration of quantum gravity effects
on the dynamics of a field. 

\subsection{Background independent dynamics}
We finish this example with the question concerning the physical content of this solution.
For the reasons we gave in the previous section, we concentrate only on the component $\Phi_+$.
Can one interpret $\Phi_+(t)$ as the position $q(t)$ of a particle evolving in the fuzzy space? 
If the answer is positive,
it is quite confusing because the position should be discrete valued whereas $\Phi_+$ takes value in the
whole real line a priori. In fact, we would like to interpret $\Phi_+(t)=Q(t) \in \mathbb R$ as the extension in 
the whole real 
line of a discrete position $q(t) \in \mathbb Z$. More precisely, we suppose that the space where the
particle evolves is one-dimensional and discrete, and then its motion should be caracterized by a $\mathbb Z$-valued
function $q(t)$. If we restore the discreteness of the time variable, then the motion of the particle should be
in fact  caracterized by a set of ordered integers $\{q(2k\ell_P),k\in \mathbb Z\}$.
To make this description more concrete, we make use of the identity satisfied by $S_+^{(1)}$ 
(\ref{identity for S})  which implies that:
\begin{eqnarray}
 Q(t)\;=\; \sum_{k=-\infty}^{+\infty} (2\ell_P k) \, K^{2k}_+(Q(t)) \;.
\end{eqnarray}
This identity makes clear that $Q(t)$ can be interpreted as a kind of continuation in the whole real line
of a set of discrete positions and $K^{2k}_+(Q(t))$ gives
the (positive or negative) weight of the discrete point $2\ell_P k$ in the evaluation of the continuous point $Q(t)$. 
Therefore, one can associate an amplitude ${\cal P}(k \vert \tau)$ to the particle 
when it is at the discrete position $Q=2\ell_P k$ and at the discrete time $t=2 \ell_P \tau$ (in Planck units) 
in the fuzzy space. This amplitude is given by:
\begin{eqnarray}\label{ampliP}
{\cal P}(k\vert\tau) \; = \; 
\frac{K^{2k}_+(Q(2\ell_P \tau))}{\sum_{j=-\infty}^{+\infty}K^{2j}_+(Q(2\ell_P \tau))} \; = \; 
K^{2k}_+(Q(2\ell_P \tau)) 
\end{eqnarray}
because the normalisation factor equals one. 
These amplitudes cannot really be interpreted as statistical weight because they can be positive or negative.
Nevertheless, they contain all the information of the dynamics of the particle in the sense that one can 
reconstruct the dynamic from these data. 
Therefore, we obtain a background independent description of the dynamics
of the particle that can be a priori anywhere at any time: its position $2k\ell_P$ at a given time 
$2\tau \ell_P$ is caracterized
by the amplitude previously defined. 
Furthermore, the amplitude is maximum around the classical trajectory, i.e. when
$Q(2\ell_P \tau)=2\ell_P k$, and gives
back the classical trajectory at the classical limit defined by $k,\tau \rightarrow \infty$, $\ell_P \rightarrow 0$ with
the products $k\ell_P$ and $\tau\ell_P$ respectively fixed to the values $t$ (classical time) and $Q$ (classical position).

\medskip

We hope to generalize this interpretation for more general (relativistic) dynamics. 
described by a (continuous) vector $Q_\mu(s)$ 
which is a function of a parameter $s$ (that can be the time component or something else). 
Indeed, there exists a relation generalizing (\ref{ampliP}) given by:
\begin{eqnarray}
Q_\mu(s) \; = \; \ell_P \sum_{j} d_j \; 
\text{tr}\left( D^j({J}_\mu) \; {K}_+^j(Q_\mu(s))\right) \;.
\end{eqnarray}
The matrix-valued function $K_+^j$ can 
be expressed in terms of special functions and its expression depends on the choice of the 
momenta functions $P_a$. Whatever the choice of $P_a$ we make, these functions admits the same classical
behavior.

Let us give an interpretation of this general formula.
A fuzzy point is parametrized by its radius fixed by the representation $j$ and its "angles" fixed by 
the magnetic numbers
$i,j \in[-I,I]$. As the fuzzy radius $R$ is a Casimir, 
it is possible to measure simultaneously the fuzzy radius and 
the fuzzy $z$ component for instance. Then, we interpret the following fonction
\begin{eqnarray}
{\cal P}(I,i\vert s) \; = \; d_j \; K^j(Q_\mu(s))^i_i
\end{eqnarray}
as the amplitude associated to a particle when it is on the sphere of radius $R=\ell_P\sqrt{I(I+1)}$ with $z=\ell_Pi$.
Thus, one would have a background independent description of the dynamics.

\section{Discussion and perspectives}
This article was mainly devoted to the study of the dynamics of a one-dimensional field in
a given non-commutative geometry.
This system is physically interpreted as a particle evolving in an Euclidean
three-dimensional quantum geometry  
which is supposed to reproduce space at the Planck scale. In a first part, we have recalled 
the basic properties of this quantum
background presenting in particular its different representations: the momentum space representation
$C(SU(2))^*$, the fuzzy space representation $\text{Mat}(\mathbb C)$ and the continuous one $C_{\ell_P}(\mathbb E^3)$. 
We have constructed the basic ingredients to define a quantum field theory on such a space: an invariant integral 
and derivative operators. Then, we write the general action for a scalar field with the requirements that the action
is local with respect to the non-commutative product and also ``invariant" by the action of the deformed symmetry algebra
$DSU(2)$.
When the field is free, solutions are similar to classical ones (i.e. solutions
of free fields equations on a classical geometry). 
Quantum gravity effects are non-trivial when one considers self-interacting fields. To illustrate this point,
we study the dynamics of a particle, instead of those of a field, in a non-linear potential. 
We show that the particle is in fact described by a couple of functions $(\Phi_+(t),\Phi_-(t))$, the first
one describes the motion of the particle and the second one the reverse motion because we have $\Phi_-(t)=\Phi_+(-t)$:
we interpret $\Phi_-(t)$ as the motion of a ``miror" particle with respect to $\Phi_+(t)$.
We find the equations of motion for $\Phi_\pm$, 
compute their solutions  at the first order (in the amplitude of the non-linear potential) 
and found  differences with classical solutions.  This is a very nice feature of our toy-model.
Let us emphasize that the quantum gravity effects are a consequence of the discretization of space-time.
Similar phenomena occur when discretizing a dynamics for numerical purposes for instance and it has been noticed
for a long time that the discretization have strong effect on the dynamics. The main novelty in our model is that
the discretization is not put by hand, on contrary it is found from fundamental principles. Furthermore, there
is a symmetry (quantum) algebra behind our construction. It would be interesting to study in great details
the effects of that discretization in the dynamics of a general field, in particular to see whether the dynamics,
when discretized according to these rules, becomes chaotic or not.

Finally, 
we propose a background independent interpretation of the dynamics of the particle defining in 
particular an amplitude associated the particle
when it is located at a given fuzzy point at a given time. 
This amplitude can be positive
or negative (so it cannot be really interpreted as a propability) and is maximal near the classical trajectory.

Nevertheless, the model is based on three dimensional Euclidean quantum gravity. What about if space-time becomes 
Lorentzian? and if space-time is four dimensional? The later question is rather difficult to answer but we can try to 
apply our technique in the LQG background. Indeed, it has been proposed a description of four-dimensional geometry
in terms of non-commutative fuzzy space \cite{CZ}. 
The former is much easier to deal with because it should be a straightforward
generalisation of our construction. However, many differences should occur due to the fact that the 
momentum space of the particle is still curved but non-compact. Therefore, the quantum
background is still expected to be non-commutative but might be
no-longer (completely) discrete. This Lorentzian regime certainly deserves to be studied in details.

\subsubsection*{Aknowlegments}
I would like to warmfully thank E. Joung and J. Mourad for their interest in the subjetc, our
numerous discussions and our fruitfull collaboration on the companion paper. 
I also want to thank A. Mouchet for very interesting and useful discussions concerning
the dynamics with a discretized time.
This work was partially supported by the ANR (BLANL06-3\_139436 LQG-2006).

\bibliographystyle{unsrt}

\end{document}